\documentclass[fleqn]{article}
\usepackage{amsmath,amssymb}
\usepackage{epsfig,graphicx}

\catcode`@=11 \@addtoreset{equation}{section} \catcode`@=12

\begin{document}
\title{\vskip-1.7cm \bf  Zero modes, gauge fixing, monodromies, $\zeta$-functions and all that}
\date{}
\author{A.O.Barvinsky and D.V.Nesterov}
\maketitle
\hspace{-8mm} {\,\,\em Theory Department, Lebedev
Physics Institute, Leninsky Prospect 53, Moscow 119991, Russia}
\begin{abstract}
We discuss various issues associated with the calculation of the reduced functional determinant of a special second order differential operator $\mbox{\boldmath${F}$}=-d^2/d\tau^2+\ddot g/g$, $\ddot g\equiv d^2g/d\tau^2$, with a generic function $g(\tau)$, subject to periodic and Dirichlet boundary conditions. These issues include the gauge-fixed path integral representation of this determinant, the monodromy method of its calculation and the combination of the heat kernel and zeta-function technique for the derivation of its period dependence. Motivations for this particular problem, coming from applications in quantum cosmology, are also briefly discussed. They include the problem of microcanonical initial conditions in cosmology driven by a conformal field theory, cosmological constant and cosmic microwave background problems.
\end{abstract}

\section{Introduction}
Several essays on calculational methods in quantum problems, we are honored to present here, are dedicated to Professor Dowker to praise his fundamental and thought-provoking contributions to quantum field theory in curved spacetime, spectral geometry and various methods of mathematical physics. His results in quantum theory with external gravitational and matter fields, pioneering contribution to the calculational method of zeta-function, treatment of zero modes, conformal anomalies and boundary terms, etc. determine the scope of methods and issues that we are going to discuss here.

The importance of these methods follows from a simple fact that successful applications on nontrivial backgrounds always present a challenge and can be accomplished in a closed analytic form only in a limited number of cases. This is equally true with regard to calculations in the quantum mechanical sector of field models even despite simplifications occurring in this sector due to its spatial homogeneity or other symmetries. A particular case of such calculations is the class of problems involving the differential operator of the form
    \begin{eqnarray}
    \mbox{\boldmath${F}$}=-\frac{d^2}{d\tau^2}+\frac{\ddot g}g,     \label{operator}.
    \end{eqnarray}
where $g=g(\tau)$ is a rather generic function of its variable $\tau$. From calculational viewpoint, the virtue of this operator is that $g(\tau)$ represents its explicit basis function -- the solution of the homogeneous equation,
    \begin{eqnarray}
    \mbox{\boldmath${F}$}g(\tau)=0,  \label{equationg}
    \end{eqnarray}
which immediately allows one to construct its second linearly independent solution
    \begin{eqnarray}
     &&\varPsi(\tau)= g(\tau)
     \int\limits_{\tau^*}^{\tau}
     \frac{dy}{g^2(y)}                 \label{Psi0}
    \end{eqnarray}
and explicitly build the Green's function of $\mbox{\boldmath${F}$}$ with appropriate boundary conditions. On the other hand, from physical viewpoint this operator is interesting because it describes long-wavelength perturbations in early Universe, including the formation of observable CMB spectra \cite{MFB,Mukhanov}, statistical ensembles in quantum cosmology \cite{PIQC}, etc. In particular, for superhorizon cosmological perturbations of small momenta $k^2\ll\ddot g/g$ their evolution operator only slightly differs from (\ref{operator}) by adding $k^2$ to its potential term, whereas in the minisuperspace sector of cosmology, corresponding to spatially constant variables, the operator has exactly the above form.

Up to an overall sign, this operator is the same both in the
Lorentzian and Euclidean signature spacetimes with the time
variables related by  the Wick rotation $\tau=it$. In the Euclidean case it plays a very important role in the calculation of the statistical sum for the microcanonical ensemble in cosmology. This ensemble realizes the concept of cosmological initial conditions by generalizing the notion of the no-boundary wavefunction of the Universe \cite{HH} to the level of a special quasi-thermal state and puts it on the basis of a consistent canonical quantization \cite{slih,why}. This concept is very promising both from the viewpoint of foundations of quantum cosmology and their applications within
the cosmological constant, inflation and dark energy problems \cite{slih,why,DGP/CFT,tunnel,PIQC}.

The recently suggested alteration in foundations of quantum cosmology -- the theory of initial conditions for the early Universe -- consists in a qualitative extension of the class of its initial quantum states. Instead of a usually accepted pure state, like the no-boundary one, it is assumed that the cosmological state can be mixed and characterized by the density matrix \cite{slih}. Under a natural and most democratic assumption of the microcanonical distribution, this density matrix and its statistical sum can be rendered the form of the Euclidean quantum gravity path integral \cite{why}. Its calculation then shows if it is dominated either by the contribution of a pure state or a mixed statistical ensemble. Thus the dilemma of pure vs mixed state, rather than being postulated, gets solved at the dynamical level according to the matter content of the model.

For models dominated by heavy massive fields this microcanonical ensemble reduces to the pure vacuum no-boundary or tunneling state \cite{tunnel}, whereas for massless conformally invariant fields the situation becomes even more interesting. In this case of the CFT driven cosmology the microcanonical ensemble incorporates a possible solution of the cosmological constant problem -- the restriction of the range of the primordial $\Lambda$ by a new quantum gravity scale which is encoded in the conformal anomaly of the underlying CFT \cite{slih,why}. Moreover, it contains a mechanism of formation of the red tilted CMB spectrum complementary (or maybe even alternative) to the conventional mechanism based on primordial vacuum fluctuations in the early inflationary Universe \cite{MukhanovChibisov}. As it was first observed in \cite{DGP/CFT} this follows from a simple fact that thermal corrections to the CMB spectrum enhance its infrared part. In connection with this, Professor Dowker might perhaps be interested to know how his pioneering results on conformal anomaly \cite{anomaly} and thermal quantum effects \cite{highT} in gravity theory underlie the foundations of quantum cosmology and, perhaps, have explicit manifestation in the most fundamental achievement of contemporary theoretical and observational cosmology -- explanation of the large scale structure of the Universe.

In this statistical theory context the operator (\ref{operator}) arises in the one-loop approximation for the cosmological statistical sum with $\tau$ playing the role of the Euclidean time, and the properties of this operator essentially differ from those of the Lorentzian dynamics. In the latter case the function $g$ is a monotonic function of time because of the monotonically growing cosmological scale factor, whereas in the Euclidean case $g(\tau)$ is periodic just as the scale factor $a(\tau)$ itself and, moreover, has zeroes at turning points of the Euclidean evolution with $\dot
a=0$, because $g(\tau)\varpropto\dot a(\tau)$. This does not lead to a singular behavior of $\mbox{\boldmath${F}$}$ because $\ddot g$ also vanishes at the zeroes of $g$ \cite{slih,PIQC}, and the potential term of (\ref{operator}) remains analytic (both $g$ and $\ddot g$ simultaneously have a {\em first-order} zero). Nevertheless, the calculation of various quantities associated with this operator becomes cumbersome due to the roots of $g(\tau)$ -- in particular, the basis function (\ref{Psi0}) becomes singular at each of these roots and cannot be extended beyond any of them. Among such quantities is the functional determinant of $\mbox{\boldmath${F}$}$ which determines the one-loop contribution to the statistical sum of the CFT driven cosmology of \cite{PIQC}. Since this operator has an obvious zero mode which is the function $g(\tau)$ itself, the functional determinant of $\mbox{\boldmath${F}$}$ should, of course, be understood as calculated on the subspace of its nonzero modes.
The focus of this paper will be a collection of issues associated with the calculation of such a {\em restricted} functional determinant. It will be denoted below by ${\rm Det^*}\mbox{\boldmath${F}$}$.

Our work largely has a methodological nature and its results in this or that context could be found in numerous papers on mathematical physics. However, their collection below portrays a rather illuminating complementarity of various methods which, on the one hand, are focused on calculation of this quantity and, on the other hand, embrace rather different fields of mathematical physics, ranging from quantization of gauge theories to monodromy method in integrable systems, heat kernel theory, spectral geometry, etc.

The calculation of ${\rm Det^*}\mbox{\boldmath${F}$}$ begins with the remark that there exist several different methods for restricted functional determinants. When the whole spectrum of the operator is known this is just the product of all non-zero eigenvalues. With the knowledge of only the zero mode, one can use the regularization technique \cite{McK-Tarlie} or contour integration method
\cite{BKK,Kirsten-McK,Kirsten-McK1} to extract the regulated
zero-mode eigenvalue from the determinant and subsequently take the regularization off. Here we use another approach to the definition of ${\rm Det^*}\mbox{\boldmath${F}$}$ based on the Faddeev-Popov gauge-fixing procedure for the path integral in quantization of gauge theories \cite{FaddeevPopov}. In Sect.2 we interpret the zero mode of $\mbox{\boldmath${F}$}$ as a generator of the gauge invariance transformation of the relevant action, so that the reduced functional determinant arises as a result gauge-fixed Gaussian path integration. This allows one to express it in terms of the Green's function of the operator $\mbox{\boldmath${F}$}$ on the subspace of its non-degeneracy. Remarkably, this Green's function follows from very simple and clear identical transformations under the path integration sign, rather than from verbose explanations one usually encounters in numerous works on the treatment of soliton or instanton zero modes.

In Sect.3 we go over to the calculation of ${\rm Det^*}\mbox{\boldmath${F}$}$ by the monodromy method. First we introduce the {\em monodromy} of the second basis function $\psi(\tau)$ of $\mbox{\boldmath${F}$}$, which is linearly independent of $g(\tau)$. In particular, we derive the answer for this monodromy for the so-called multiple nodes case, when the function $g(\tau)$ within its period range has an arbitrarily high even number $2k$ of roots, $g(\tau_i)=0$, $i=1,2,...,2k$.\footnote{Since a periodic function has only an even number of roots within its period, we will call the case of their lowest nonvanishing number, $2k=2$, the {\em single-node} one. This is the case of the CFT driven cosmology whose statistical sum as a function of the primordial cosmological constant is dominated by the countable set of instantons having $k$ oscillations, $k=1,2,...$, of the cosmological scale factor $a(\tau)$ during the Euclidean time period \cite{slih,why} -- the so-called garlands which carry the multi-node zero mode $g(\tau)\propto\dot a$.} The monodromy is presented as an additive sum of contributions of segments of the time variable, $\tau_{i-1}\leq\tau\leq\tau_i$, separating various pairs of neighboring roots. Each contribution is given by a closed integral expression in terms of $g(\tau)$ on an underlying segment. Then, by the variational method for the functional determinant, we express ${\rm Det^*}\mbox{\boldmath${F}$}$ in terms of this monodromy. This actually reproduces the known monodromy formula of McKane and Tarlie for the restricted functional determinant with periodic boundary conditions \cite{McK-Tarlie,Kirsten-McK,Kirsten-McK1}, but leaves undefined its overall normalization coefficient which is functionally independent of $g(\tau)$, but can be a function of $T$ -- the period of the time range.

Sect.4 is devoted to the calculation of this coefficient, and this is a place where various methods -- WKB approximation, zeta-function technique, heat kernel method, spectral geometry and effect of boundaries get efficiently intertwined and complement each other. We begin with the case of the nondegenerate operator (\ref{operator}) subject to Dirichlet boundary conditions on a finite segment of time $[\tau_-,\tau_+]$, $\tau_+-\tau_-=T$ (with a rootless function $g(\tau)$). It gives the preexponential factor of the time evolution operator between $\tau_-$ and $\tau_+$, whose dependence on $T$ can be obtained by the combination of the $\zeta(s)$-function method \cite{zeta} and conformal rescaling incorporating a particular value of the zeta-function, $\zeta(0)$, responsible for the conformal anomaly of the theory \cite{anomaly}. The latter is obtained from the spectral geometry method and the proper time expansion of the heat kernel \cite{DW,Gilkey-Seely,PhysRep,Avramidi,Vassilevich} in the one-dimensional spacetime with two boundaries at $\tau_\pm$. On the other hand, the prefactor of this evolution operator with the time-dependent coefficient inclusive is given by the Pauli-van Vleck-Morette formula \cite{PvVM} for the unitary evolution operator (or its Euclidean version \cite{reduc}). Comparison of these two results gives a correct answer for the normalization coefficient. Professor Dowker might perhaps be amused to see how this simple effect of boundary terms in the heat kernel expansion recovers a correct time dependence of the evolution operator dictated by the Schrodinger equation, the situation immeasurably more simple than the one he considered in his pioneering work on boundary effects of quantum polarization in curved spacetime \cite{Dowkerboundaries,Dowkerboundaries1}. We accomplish Sect. 4 by applying the same zeta-function method to our multi-node case of the operator (\ref{operator}) subject to periodic boundary conditions and prove the $T$-independent monodromy algorithm for its reduced functional determinant.  Sect. 5 contains conclusions and the discussion of possible applications of the above formalism.

\section{Zero modes and Faddeev-Popov gauge fixing for reduced functional determinants}
When the operator $\mbox{\boldmath${F}$}$ is nodegenerate its functional determinant can of course be determined by the Gaussian functional integral
    \begin{eqnarray}
    &&({\rm Det}_{\cal D}{\mbox{\boldmath${F}$}})^{-1/2}
    ={\rm const}\times\int D\varphi\;
    \exp\Big\{-S[\,\varphi\,]\,\Big\},         \label{Gaussian}
    \end{eqnarray}
with the quadratic action
    \begin{eqnarray}
    S[\,\varphi\,]=\frac12\int_{\cal D}
    d\tau\,\varphi(\tau)\mbox{\boldmath${F}$}
    \varphi(\tau)=\frac12\int_{\cal D}
    d\tau\,\left(\dot\varphi^2
    +\frac{\ddot g}g\,\varphi^2\right),         \label{action}
    \end{eqnarray}
where the domain of integration ${\cal D}$ and boundary conditions on integration variable $\varphi(\tau)$ -- one-dimensional field -- are determined by the class of fields on which the determinant ${\rm Det}_{\cal D}{\mbox{\boldmath${F}$}}$ is defined (in what follows the label ${\cal D}$ will signify both the domain of $\tau$ and relevant boundary conditions for which the operator $\mbox{\boldmath${F}$}$ is Hermitian).

The zero mode $g(\tau)$ of (\ref{operator}), provided it satisfies these boundary conditions, arises as the
generator of the global gauge invariance of the action (\ref{action}) under the transformation with a constant $\varepsilon$,
    \begin{eqnarray}
    &&\delta^\varepsilon\varphi
    =R(\tau)\,\varepsilon,          \label{gt}\\
    &&R(\tau)=\frac{g(\tau)}{||g||},\quad
    ||g||^2=\int_{\cal D} d\tau\,g^2(\tau).       \label{generator}
    \end{eqnarray}
Therefore, for a degenerate operator with the single zero mode the path integral representation of its functional determinant (\ref{Gaussian}) can be handled by means of the well-known Faddeev-Popov gauge fixing procedure \cite{FaddeevPopov}.  It consists of imposing the gauge $\chi[\,\varphi\,]=0$ and inserting in  the path integral the relevant Faddeev-Popov factor. This  gauge condition $\chi[\,\varphi\,]$ and the Faddeev-Popov ghost factor $Q/||g||$ can be chosen in the form
    \begin{eqnarray}
    &&\chi[\,\varphi\,]
    =\int_{\cal D} d\tau\,k(\tau)\,\varphi(\tau),
    \quad \delta^\varepsilon\chi
    =\frac{Q}{||g||}\,\varepsilon,          \label{gauge0}\\
    &&Q\equiv\int_{\cal D} d\tau\,k(\tau)\,g(\tau),       \label{Q}
    \end{eqnarray}
where $k(\tau)$ is a gauge fixing function and the generator (\ref{generator}) is normalized to unity with respect to $L^2$ inner product on $\cal D$. Thus, integration over $\varphi$ takes the form of the Gaussian functional integral with the delta-function type gauge
    \begin{eqnarray}
    &&({\rm Det}_{\cal D}^*{\mbox{\boldmath${F}$}})^{-1/2}
    ={\rm const}\times\int D\varphi\;
    \delta\Big(\chi[\,\varphi\,]\Big)\,\frac{Q}{||g||}
    \exp\Big\{-S[\,\varphi\,]\,\Big\},         \label{I}
    \end{eqnarray}
and serves as the definition of the restricted functional determinant of $\mbox{\boldmath${F}$}$. This definition is in fact independent of the choice of gauge by the usual gauge independence mechanism for the Faddeev-Popov integral. In particular, enforcing the gauge $\chi=0$ means that the field $\varphi$ is functionally orthogonal to the gauge fixing function $k(\tau)$ in the $L^2$ metric on $\cal D$, and the above definition is independent of the choice of this gauge fixing function.

The normalization of the generator (\ref{generator}) has the following explanation. In local gauge theories the Faddeev-Popov path integral is not invariant under arbitrary rescalings of gauge generators. Their normalization is always implicitly fixed by the requirement of locality and the unit coefficient of the time-derivative term in the gauge transformation of Lagrange multipliers, $R\varepsilon\sim 1\times\dot\varepsilon+...$, (which follows from the canonical quantization underlying the Hamiltonian version of the Faddeev-Popov path integral). For the global symmetry of (\ref{action}) we do not have a counterpart in canonical formalism, and such a founding principle as canonical quantization does not seem to be available. Therefore, we choose this normalization with respect to $L^2$ unit norm corresponding to the canonical normalization of the variable $\varphi$ in (\ref{action}). From the viewpoint of the definition of ${\rm Det}_{\cal D}^*\mbox{\boldmath${F}$}$ as the product of operator eigenvalues, this corresponds to the omission of a zero eigenvalue of $\mbox{\boldmath${F}$}$,
    \begin{eqnarray}
    &&{\rm Det}_{\cal D}^*\,
    {\mbox{\boldmath${F}$}}
    =\prod_{\lambda\neq 0}\lambda,   \label{prodlambda}\\
    &&\mbox{\boldmath${F}$}
    \varphi_\lambda(\tau)
    =\lambda\,\varphi_\lambda(\tau),
    \quad \tau\in{\cal D}.            \label{eigenproblem}
    \end{eqnarray}
This follows from the orthogonal decomposition of the integration variable $\varphi(\tau)$ in the series of eigenfunctions $\varphi_\lambda(\tau)$ satisfying
    \begin{eqnarray}
    \int_{\cal D} d\tau\,\varphi_\lambda(\tau)\,
    \varphi_{\lambda'}(\tau)
    =\delta_{\lambda\lambda'}.
    \end{eqnarray}

Representing the delta function of the gauge condition in (\ref{I})
via the integral over the Lagrangian multiplier $\pi$ we get the
Gaussian path integral over the periodic function $\varphi(\tau)$
and the numerical variable $\pi$,
    \begin{eqnarray}
    &&({\rm Det}_{\cal D}^*{\mbox{\boldmath${F}$}})^{-1/2}
    ={\rm const}\times Q\,||g||^{-1}\int D\varphi\,d\pi\,\exp\Big(-
    S_{\rm eff}[\,\varphi(\tau),\pi\,]\,\Big)\nonumber\\
    &&\qquad\qquad\qquad={\rm const}\times Q\,||g||^{-1}\Big({\rm Det}\,
    \mathbb{F}\Big)^{-1/2}.                         \label{I2}
    \end{eqnarray}
Here $S_{\rm eff}[\,\varphi(\tau),\pi\,]$ is the effective action of
these variables and $\mathbb{F}$ is the matrix valued Hessian of
this action with respect to $\varPhi=(\varphi(\tau),\pi)$,
    \begin{eqnarray}
    &&S_{\rm eff}[\,\varphi(\tau),\pi\,]=S[\,\varphi\,]-i\pi \int_{\cal D}
    d\tau\,k\varphi,\\
    &&\mathbb{F}=
    \frac{\delta^2S_{\rm eff}}
    {\delta\varPhi\, \delta\varPhi'}=
    \left[\,\begin{array}{cc} \;{\mbox{\boldmath${F}$}}\,
    \delta(\tau,\tau')&\,\,\, -i k(\tau)\,\\
    &\\
    -i k(\tau')&0\end{array}\,\right]     \label{matrixF}
    \end{eqnarray}
(note the position of time entries associated with the variables
$\varPhi=(\varphi(\tau),\pi)$ and $\varPhi'=(\varphi(\tau'),\pi)$).

The dependence of this determinant on $g(\tau)$ and $k(\tau)$ can be found from its variation with respect to these functions. From (\ref{I2}) we have
    \begin{eqnarray}
    &&\delta\ln\Big({\rm Det_{\cal D}^*}\,
    \mbox{\boldmath${F}$}\Big)
    =-2\,\delta\ln Q+2\,\delta\ln ||g||
    +{\rm Tr}\,\Big(\delta\mathbb{F}\,
    \mathbb{G}\Big),                         \label{var1}
    \end{eqnarray}
where $\mathbb{G}$ is the Green's function of $\mathbb{F}$,
$\mathbb{F}\,\mathbb{G}=\mathbb{I}$ and the functional trace of any matrix with the block-structure of (\ref{matrixF}) is defined as
    \begin{eqnarray}
    &&{\rm Tr}
    \left[\,\,\begin{array}{cc}A(\tau,\tau')\,&\,\,\,
    B(\tau)\,\\
    B(\tau')&a\end{array}
    \,\right]=\int_{\cal D} d\tau\,A(\tau,\tau)+a\,.  \nonumber
    \end{eqnarray}

The block structure of the
matrix Green's function $\mathbb{G}$ has the form
    \begin{eqnarray}
    \mathbb{G}=
    \left[\,\,\begin{array}{cc}G(\tau,\tau')\,&\,\,\,
    {\displaystyle \frac{\textstyle i g(\tau)}Q}\,\\
    {\displaystyle \frac{\textstyle i g(\tau')}Q}
    &0\end{array}
    \,\right],                                    \label{matrixG}
    \end{eqnarray}
where the Green's function $G(\tau,\tau')$ in the diagonal block
satisfies the system of equations
    \begin{eqnarray}
    &&{\mbox{\boldmath$F$}}\,
    G(\tau,\tau')=\delta(\tau,\tau')
    -\frac{k(\tau)\,g(\tau')}Q,            \label{Gequation}\\
    &&\int_{\cal D} d\tau\,g(\tau)\,G(\tau,\tau')
    =0,                                           \label{Ggauge}
    \end{eqnarray}
which uniquely fix it. The second equation imposes the needed gauge,
whereas the right hand side of the first equation implies that
$G(\tau,\tau')$ is the inverse of the operator $F$ on the subspace
orthogonal to its zero mode.

The trace of the functional block-structure matrix in (\ref{var1}) corresponding to the variation of $g(\tau)$ reads
    \begin{eqnarray}
     &&{\rm Tr}\,\Big(\delta_g\mathbb{F}\,
     \mathbb{G}\Big)=
     \mathrm{Tr}\Big(\delta\mbox{\boldmath$F$}\,
     G(\tau,\tau')\Big)\equiv\int_{\cal D} d\tau\,
    \delta\mbox{\boldmath$F$}\,
    G(\tau,\tau')\Big|_{\,\tau'=\tau}.      \label{deltagFG}
    \end{eqnarray}
A similar variation of the gauge-fixing function gives a vanishing answer $\delta_k\ln\big({\rm Det_{\cal D}^*}\, \mbox{\boldmath$F$}\big)=0$
as, of course, it should be in view of the gauge independent nature
of the Faddeev-Popov path integral.\footnote{In the works involving the treatment of soliton and instanton zero modes it is implicitly assumed that the gauge-fixing function coincides with the zero mode itself, $k(\tau)=g(\tau)$, which considerably simplifies the formalism, but makes it less flexible.} This guarantees the uniqueness of the definition of the reduced determinant ${\rm Det_{\cal D}^*}\, \mbox{\boldmath$F$}$.

\section{Periodic boundary conditions and the monodromy method}

Here we consider periodic boundary conditions for the operator (\ref{operator}) which is defined on a circle range of the time variable ${\cal D}=S^1$ having the circumferance $T$. It is parameterized by $\tau$
    \begin{eqnarray}
    \tau_0<\tau<\tau_0+T,        \label{circle}
    \end{eqnarray}
with the points $\tau_0$ and $\tau_0+T$ being identified, so that
integration over this range will be denoted by
    \begin{eqnarray}
    \int_{\cal D} d\tau\,(...)\equiv\oint d\tau\,(...).
    \end{eqnarray}

This range can be infinitely extended to the whole axis $-\infty<\tau<\infty$, multiple covering of $S^1$, on which the function $g(\tau)$ and, consequently, the operator $\mbox{\boldmath${F}$}$ are periodic with the period $T$,
    \begin{eqnarray}
    g(\tau+T)=g(\tau).        \label{periodicg}
    \end{eqnarray}

The problem of major interest here will be the so-called multi-node case, motivated as it was mentioned in Introduction by applications in cosmology, when the periodic function $g(\tau)$ is oscillating and has within its period $2k$ simple roots
    \begin{eqnarray}
    &&\tau_0<\tau_1<\tau_2<...
    \tau_{2k}=\tau_0+T,          \label{roots}\\
    &&g(\tau_i)=0,\quad
    \dot g(\tau_i)\neq 0,      \label{simpleroots}\\
    &&\ddot g(\tau_i)=0.       \label{ddotg}
    \end{eqnarray}
For simplicity we assume that one of them coincides with the final (or starting) point of this period. Another important assumption is that the second order derivative of this function at its roots is vanishing, which will be important for analyticity properties of our formalism.

Another important property of the operator $\mbox{\boldmath${F}$}$ is its Wronskian relation. For any two functions $\varphi_1$ and $\varphi_2$ this operator determines their Wronskian $W[\varphi_1,\varphi_2]\equiv \varphi_1\dot\varphi_2-\dot\varphi_1\varphi_2$ which enters the relation
    \begin{eqnarray}
    &&\int\limits_{\tau_-}^{\tau_+} d\tau \,\varphi_1 \overrightarrow{\mbox{\boldmath$F$}} \varphi_2
    =\int\limits_{\tau_-}^{\tau_+} d\tau \,\varphi_1 \overleftarrow{\mbox{\boldmath$F$}}
    \varphi_2 - W[\varphi_1,\varphi_2]\,
    \Big|^{\;\tau_+}_{\;\tau_-}.        \label{Wronskianrelation}
    \end{eqnarray}
Arrows here denote the direction of action of the operator $\mbox{\boldmath$F$}$, i. e. $\varphi_1\! \overleftarrow{\mbox{\boldmath$F$}}=(\mbox{\boldmath$F$}\varphi_1)$, and the Wronskians appear as total derivative terms generated by integration by parts of the derivatives in $\mbox{\boldmath$F$}$. When both $\varphi_1$ and $\varphi_2$ satisfy a homogeneous equation with the operator $\mbox{\boldmath${F}$}$, their Wronskian turns out to be constant. Also the vanishing Wronskian implies linear dependence of these solutions.

\subsection{Monodromy for the multi-node case}

Let us now consider the solution of the homogeneous equation $\psi(\tau)$ normalized by a unit value of its Wronskian with $g$
    \begin{eqnarray}
    &&\mbox{\boldmath${F}$}\psi(\tau)=0,
    \quad W[g,\psi]=1.                  \label{equationpsi}
    \end{eqnarray}
Together with $g(\tau)$ this solution forms a set of linearly independent basis functions of $\mbox{\boldmath$F$}$. However, in contrast to $g(\tau)$ the basis function is not periodic, because we assume that the operator (\ref{operator}) has only one periodic zero mode smoothly defined on a circle (\ref{circle}). On the other hand,
when considered on the full axis of $\tau$, due to periodicity of $g(\tau)$ this operator is also periodic $\mbox{\boldmath$F$}(\tau+T)=\mbox{\boldmath$F$}(\tau)$. Therefore $\psi(\tau+T)$ is also a solution of the equation $\mbox{\boldmath$F$}(\tau)\psi(\tau+T)=0$, and consequently it can be decomposed into a linear combination of the original two basis functions with constant coefficients
    \begin{eqnarray}
    \psi(\tau+T)=\psi(\tau)+\Delta\, g(\tau).     \label{Monpsi}
    \end{eqnarray}
The unit coefficient in the first term follows from the conservation in time of the Wronskian of any two solutions of the equation (\ref{equationpsi}), periodicity of $g(\tau)$ and an obvious fact that $W[g,g]=0$ and
    \begin{eqnarray}
    1=W[g(\tau+T),\psi(\tau+T)]=W[g(\tau),\psi(\tau) +\Delta\, g(\tau)].
    \end{eqnarray}
The coefficient $\Delta$ in the second term of (\ref{Monpsi}) is nontrivial -- this is the {\em monodromy} of $\psi(\tau)$ which will play a central role in the construction of the determinant.

The function $\psi(\tau)$ can be composed of the set of functions $\varPsi_i(\tau)$ defined by (\ref{Psi0}) on various segments of $\tau$-range connecting the pairs of neighboring roots of $g(\tau)$\footnote{For the extended range (\ref{circle1}) the missing $\varPsi_0(\tau)$ can be defined by identifying $\tau_{-1}$ with $\tau_{2k-1}-T$ and choosing some $\tau_0^*$ in $\tau_{-1}<\tau_0^*<\tau_0$.}
    \begin{eqnarray}
     &&\varPsi_{i}(\tau)= g(\tau)
     \int_{\tau_{i}^*}^{\tau}
     \frac{dy}{g^2(y)},\;\quad
     \tau_{i-1}<\tau,\tau_{i}^*<\tau_i\;,
     \quad i=1,...,2k                         \label{Psis}
    \end{eqnarray}
Here $\tau_i^*$ are the auxiliary points arbitrarily chosen in the same segments, and all these solutions are normalized by the unit Wronskian with $g(\tau)$, $W[g,\varPsi_i]=1$. The main property of these functions $\varPsi_i(\tau)$ is that each of them is defined in the $i$-th segment of the full period of $\tau$ where the integral (\ref{Psis}) is convergent because the roots of
$g(\tau)$ do not occur in the integration range. Its limits are well defined also at the boundaries of this segment,
    \begin{eqnarray}
    \varPsi_i(\tau_{i-1})=
    -\frac1{\dot{g}(\tau_{i-1})},\quad
    \varPsi_i(\tau_i)
    =-\frac1{\dot{g}(\tau_i)},    \label{psilimits}
    \end{eqnarray}
because the factor $g(\tau)$ tending to zero compensates for the
divergence of the integral at $\tau\to\tau_i-0$ and $\tau\to\tau_{i-1}+0$.

For an arbitrary choice of auxiliary points  $\tau_i^*$ in (\ref{Psis}) the composite function
    \begin{eqnarray}
    &&\psi(\tau)=\varPsi_i(\tau),\;
    \qquad \tau_{i-1}\leq\tau\leq\tau_i,     \label{psi}
    \end{eqnarray}
will be continuous in view of (\ref{psilimits}), but the continuity of its derivative will generally be broken, because generally the equality $\dot\varPsi_i(\tau_i-0)=\dot\varPsi_{i+1}(\tau_i+0)$ is not satisfied. However, this equality for $i=1,2,...,2k-1$ can be enforced by a special choice of these auxiliary points $\tau_i^*$, becoming the equation for their determination. The solution for $\tau_i^*$ is unique, always exists and belongs to the corresponding segment $\tau_{i-1}<\tau_i^*<\tau_i$.\footnote{Indeed, the quantity $d\dot\varPsi_i(\tau_i)/d\tau_i^*=-\dot g(\tau_i)/g^2(\tau_i^*)$ is a sign definite function of $\tau_i^*$ nowhere vanishing on this segment, its absolute value quadratically divergent to $\infty$ at its boundaries. This in its turn means that $\dot\varPsi_i(\tau_i)$ is a monotonic function of $\tau_i^*$ which also ranges between $-\infty$ and $+\infty$ and therefore guarantees the unique solution for $\tau_i^*$ on this segment.} On the other hand, the continuity of the derivative of $\psi(\tau)$ cannot be attained at all roots of $g(\tau)$, $i=1,2,...2k$, because it would correspond to the existence of the second zero mode periodic on the circle, which is ruled out by construction. Therefore, the second basis function of $\mbox{\boldmath$F$}$ is not periodic on the circle, but in view of periodicity of the operator it satisfies the fundamental monodromy property (\ref{Monpsi}). In the next subsection we construct the periodic Green's function of $\mbox{\boldmath$F$}$ in terms of this monodromy parameter $\Delta$, whereas here we give in a closed form the analytic expression for $\Delta$ as a functional of $g(\tau)$.

From the definition of the monodromy parameter (\ref{Monpsi}) it follows that in the limit $\tau\to\tau_0$, $\tau+T\to\tau_{2k}$,
    \begin{eqnarray}
    \Delta=\frac{\dot\psi(\tau_{2k})-\dot\psi(\tau_0)}{\dot g(\tau_0)}=\left(\frac{\dot\psi(\tau_{2k})}{\dot g(\tau_{2k})}-
    \frac{\dot\psi(\tau_{2k-1})}{\dot g(\tau_{2k-1})}\right)+...+\left(
    \frac{\dot\psi(\tau_1)}{\dot g(\tau_1)}-
    \frac{\dot\psi(\tau_0)}{\dot g(\tau_0)}\right),
    \end{eqnarray}
where we took into account that $\dot g(\tau_{2k})=\dot g(\tau_0)$.
Then the monodromy reads as the additive sum of contributions of pairs of neighboring roots of $g(\tau)$ \cite{DetCFTQC},
    \begin{eqnarray}
    &&\Delta=\sum\limits_{i=1}^{2k}
    \varDelta_i,                     \label{Delta}\\
    &&\varDelta_i=
    \frac{\dot\varPsi_i(\tau_i)}{\dot g(\tau_i)}
    -\frac{\dot\varPsi_i(\tau_{i-1})}
    {\dot g(\tau_{i-1})}
    =-\Big(\varPsi_i(\tau_i)\,
    \dot\varPsi_i(\tau_i)
    -\varPsi_i(\tau_{i-1})\,
    \dot\varPsi_i(\tau_{i-1})\Big).                 \label{Deltai}
    \end{eqnarray}

Because of (\ref{ddotg}) the functions $\varPsi_i(\tau)$ are differentiable in these limits, and all the quantities which enter the algorithm (\ref{Deltai}) are well defined. In particular, for any such time segment $[\tau_{i-1},\tau_i]\equiv[\tau_-,\tau_+]$ the derivatives of $\varPsi(\tau)$ at its boundaries are given by the convergent integral
    \begin{eqnarray}
    \dot\varPsi(\tau_\pm)=
    \int\limits_{\tau^*}^{\tau_\pm} dy\,
    \frac{\dot g(\tau_\pm)
    -\dot g(y)}{g^2(y)}+\frac1{g(\tau^*)}.    \label{derivative}
    \end{eqnarray}
Note that the integrand here is finite at $y\to\tau_\pm$ because of $\ddot g(\tau_\pm)=0$. These properties of $\varPsi_i(\tau)$ guarantee that the obtained result is independent of the choice of the auxiliary point $\tau_i^*$ for each $\Delta_i$, and the monodromy (\ref{Delta}) is uniquely defined.

It is important that unlike in the construction of the function $\psi(\tau)$ which has to be smooth on $S^1$ at all roots $\tau_i$ except $\tau_0$ (the property that was attained above by a special choice of the auxiliary points $\tau_i^*$), the derivatives of neighboring functions $\varPsi_i(\tau)$ in (\ref{Delta})-(\ref{Deltai}) should not necessarily be matched at these junction points. This is because the partial contributions $\Delta_i$ to the overall monodromy $\Delta$ are individually independent of  $\tau_i^*$,
    \begin{eqnarray}
    \frac{d\varDelta_i}{d\tau_i^*}=0,
    \end{eqnarray}
which can be easily verified by using a simple relation
$d\dot\varPsi_i(\tau)/d\tau_i^*=-\dot g(\tau)/g^2(\tau_i^*)$. Thus, the monodromy is uniquely defined and independent of the choice of the auxiliary points $\tau_i^*$ necessarily entering the definition of functions $\varPsi_i(\tau)$ in Eq.(\ref{Psis}).

\subsection{Periodic Green's function and the variation of the determinant}
For the calculation of the variation (\ref{deltagFG}) above we need the Green's function of the problem (\ref{Gequation})-(\ref{Ggauge}) which should be periodic on the circle (\ref{circle}). To achieve this property we will slightly extend the circle domain to the left of the point $\tau_0$
    \begin{eqnarray}
    \tau_0-\varepsilon<\tau<\tau_0+T,
    \quad\varepsilon>0,          \label{circle1}
    \end{eqnarray}
with an arbitrarily small positive $\varepsilon$ and demand that the monodromy of $G(\tau,\tau')$ is vanishing for this small $\varepsilon$-range of $\tau$ near $\tau_0$
    \begin{eqnarray}
    G(\tau+T,\tau')-G(\tau,\tau')=0, \quad
    \tau_0-\varepsilon<\tau<\tau_0.       \label{MonodromyG}
    \end{eqnarray}

The ansatz for $G(\tau,\tau')$ can be as usual built with the aid of two linearly independent basis functions of the operator. One basis function coincides with the periodic zero mode $g(\tau)$ and another one is given by the function $\psi(\tau)$ built above. Thus it can be represented as a sum of the particular solution of the inhomogeneous equation (\ref{Gequation}) and the bilinear combination of $g(\tau)$ and $\psi(\tau)$ with the coefficients providing the periodicity property (\ref{MonodromyG}). As shown in \cite{DetCFTQC} it reads
    \begin{eqnarray}
     &&G(\tau,\tau') = G_F(\tau,\tau') + \frac1Q\, \Omega(\tau,\tau')+\alpha\, H_{\psi\psi}(\tau,\tau')
     + \beta\,H_{\psi g}(\tau,\tau')
     +\gamma\,H_{gg}(\tau,\tau')\;,    \label{G1}
    \end{eqnarray}
where
    \begin{eqnarray}
     &&G_F(\tau,\tau')
     \equiv\frac12\big(g(\tau)\psi(\tau')
     -\psi(\tau)g(\tau')\big)\;\theta(\tau -\tau')
     \;+\;\frac12\big(\psi(\tau)g(\tau')
     -g(\tau)\psi(\tau')\big)\;
     \theta(\tau'-\tau\! )\;,                    \label{G_F}\\
     &&\Omega(\tau,\tau') \,\equiv\,
     \omega(\tau)\,g(\tau')
     +g(\tau)\,\omega(\tau')\;,                 \label{G_Omega}\\
     &&H_{\psi\psi}(\tau,\tau')\,
     \equiv\,\psi(\tau)\psi(\tau')\;,            \label{H1}\\
     &&H_{\psi g}(\tau,\tau') \,\equiv\,
     \psi(\tau)\,g(\tau')+
     g(\tau)\,\psi(\tau')\;,                    \label{H2}\\
     &&H_{gg}(\tau,\tau') \,
     \equiv\, g(\tau)\,g(\tau')\;,              \label{H3}
    \end{eqnarray}
and the function $\omega(\tau)$ is defined by
    \begin{eqnarray}
     &&\omega(\tau)=\psi (\tau)
     \int\limits_{\tau_\star}^{\tau} dy \,g(y)k(y)
     - g(\tau) \int\limits_{\tau_{\ast}}^{\tau}
     dy \,\psi(y)k(y)           \label{def_omega_ast_st}
    \end{eqnarray}
with an arbitrary $\tau_*$. The first term of (\ref{G1}) generates the delta-function in the right hand side of the equation (\ref{Gequation}), the second term $\Omega(\tau,\tau')/Q$ gives $-k(\tau)\,g(\tau')/Q$, while $H_{\psi\psi}(\tau,\tau')$, $H_{\psi g}(\tau,\tau')$ and $H_{gg}(\tau,\tau')$ represent symmetric solutions of the homogeneous equation with coefficients $\alpha$, $\beta$ and $\gamma$ which are fixed by the periodicity condition and the gauge condition (\ref{Ggauge}),
    \begin{eqnarray}
     &&\alpha = - \frac1{\Delta}\,,\qquad
     \beta = -\frac12 \,+\, \frac1{\Delta \,Q}\!\int\limits_{\tau_{*}}^{\tau_{*}+T}
     \!\!dy\, \psi(y)\, k(y),                    \label{ab}\\
    &&\gamma = \frac1Q \int\limits_{\tau_{*}}^{\tau_{*}+T} \!\!dy\, \psi(y)\, k(y)-\frac1{Q^2 \Delta} \left(\int\limits_{\tau_{*}}^{\tau_{*}+T} \!\!dy\, \psi(y)\, k(y)\right)^2 - \frac1{Q^2}\int\limits_{\tau_{*}}^{\tau_{*}+T} \!\!dy\, \omega(y)\, k(y).
    \end{eqnarray}

The calculation of the variational term (\ref{deltagFG}) is based on integration by parts and a systematic use of the Wronskian relation (\ref{Wronskianrelation}) together with equations for $g$ and $\psi$. The result of this calculation is presented in much detail in \cite{DetCFTQC} and reads
    \begin{eqnarray}
    \delta\ln\Big({\rm Det_*}\,
    \mbox{\boldmath${F}$}\Big)
    =2\,\delta\ln ||g||+\delta\ln\Delta.
    \end{eqnarray}
It finally gives the explicit answer for ${\rm Det_*}\,\mbox{\boldmath${F}$}$
    \begin{eqnarray}
    {\rm Det}_{S^1}^*\,
    \mbox{\boldmath${F}$}
    =C(T)\times
    \Delta\oint d\tau\,g^2(\tau).    \label{det1}
    \end{eqnarray}

In fact, this is the McKane-Tarlie formula (Eq.(5.2) of \cite{McK-Tarlie}) obtained by the regularization and contour integration methods \cite{McK-Tarlie,Kirsten-McK,Kirsten-McK1} based on the earlier work of Forman \cite{Forman} for a generic second order differential operator. We have reproduced this formula by the variational method for functional determinants. Beyond this, the structure of the operator (\ref{operator}) makes the problem exactly solvable and gives the monodromy (\ref{Monpsi}) in quadratures as an explicit functional of $g(\tau)$.

Final comment concerns the overall normalization in (\ref{det1}). The formula of McKane-Tarlie \cite{McK-Tarlie,Kirsten-McK,Kirsten-McK1} or the variational method which we use below, in principle, give only the ratio of determinants for two different operators with different functions $g$, whereas each determinant contains an infinite numerical factor generated by UV divergent product of eigenvalues of $\mbox{\boldmath${F}$}$. This factor is independent of $g(\tau)$ but depends on the UV regularization and can be a function of the period $T$ -- the only remaining free parameter of the problem. In the next section we derive it by the Dowker $\zeta$-function method of \cite{zeta}.

\section{$\zeta$-function method and zero modes}

\subsection{Dirichlet problem and the effect of boundaries}
We begin the consideration of the $\zeta$-function method with the case of the functional determinant subject to Dirichlet boundary conditions on the initial and final points $\tau_\pm$ of the time interval of the length $T$
    \begin{eqnarray}
    {\cal D}=[\,\tau_-,\tau_+\,], \quad \tau_+-\tau_-=T.
    \end{eqnarray}
Here the focus of our attention will not be zero modes of the operator, but rather -- a simple but illuminating effect of boundaries, which in the quantum mechanical (that is in one-dimensional) case is very universal because it applies to operators of a general form. Thus, we assume that the operator (\ref{operator}) is nondegenerate, and its function $g(\tau)$ nowhere on $\cal D$ equals zero. Therefore, it is not a zero mode of the Dirichlet problem because $g(\tau_\pm)\equiv g_\pm\neq 0$.

The Dirichlet problem arises when one considers a semiclassical approximation for the kernel of the unitary evolution operator. This kernel, which is given by the path integral, on the one hand reduces in the subleading order of this expansion to the one-loop result
    \begin{eqnarray}
    &&\int\limits_{\varphi(\tau_\pm)=\varphi_\pm} \!\!\!\!D\varphi\;
    \exp\Big\{-S[\,\varphi\,]\,\Big\}={\rm const}\times\Big({\rm Det}_D{\mbox{\boldmath${F}$}}\Big)^{-1/2}
    \exp\Big\{-{\mbox{\boldmath${S}$}}\,\Big\},
    \end{eqnarray}
where ${\mbox{\boldmath${S}$}}$ is the on-shell action (\ref{action}) calculated on the solution of classical equations of motion $\varphi(\tau)=\varphi(\tau|\varphi_\pm)$ interpolating between initial and final configurations $\varphi_\pm$ at $\tau_\pm$ (the principle Hamilton function)
    \begin{eqnarray}
    &&{\mbox{\boldmath${S}$}}\equiv {\mbox{\boldmath${S}$}}(\tau_\pm,\varphi_\pm)
    =S[\,\varphi\,]\,\big|_{\;\varphi(\tau)=\varphi(\tau|\varphi_\pm)}
    \end{eqnarray}
and ${\rm Det}_D{\mbox{\boldmath${F}$}}$ is the the functional determinant subject to Dirichlet boundary conditions on $[\,\tau_-,\tau_+\,]$.

On the other hand, it is given by the Pauli-van Vleck-Morette formula as the semiclassical solution of the Schroedinger equation \cite{PvVM,reduc}\footnote{To be more precise, the Euclidean version of this equation corresponding to the imaginary time $\tau$.} corresponding to the action (\ref{action})
    \begin{eqnarray}
    &&\!\!\!\int\limits_{\varphi(\tau_\pm)=\varphi_\pm} \!\!\!\!D\varphi\;
    \exp\Big\{-S[\,\varphi\,]\,\Big\}=
    \left(-\frac1{2\pi}\frac{\partial^2{\mbox{\boldmath${S}$}}}
    {\partial\varphi_+\partial\varphi_-}\right)^{1/2}
    \exp\Big\{-{\mbox{\boldmath${S}$}}\,\Big\}.
    \end{eqnarray}
This implies the following equality between the functional determinant and the Pauli-van Vleck prefactor
    \begin{eqnarray}
    &&{\rm Det}_D{\mbox{\boldmath${F}$}}={\rm const}\times
    \left(\frac{\partial^2{\mbox{\boldmath${S}$}}}
    {\partial\varphi_+\partial\varphi_-}\right)^{-1} \label{PvVM}
    \end{eqnarray}
with a constant proportionality coefficient, which is independent of $\tau_\pm$ because the time evolution Schroedinger equation does not admit any flexibility in time-dependent normalization of its solution. Below we prove this fact by using the $\zeta$-function calculation of the left-hand side of this relation and comparing it with the explicit time dependence of the right-hand side -- the exercise illuminating the role of boundary terms.

First we find the dependence on time of the expression in the right-hand side of (\ref{PvVM}). For this we note that $\partial{\mbox{\boldmath${S}$}}/\partial\varphi_+$ is the canonical momentum conjugated to the Lagrangian variable $\varphi(\tau)$ at $\tau_+$ or $\dot\varphi(\tau_+)$. This allows one to write a set of relations
    \begin{eqnarray}
    &&\frac{\partial{\mbox{\boldmath${S}$}}}{\partial\varphi_+}=
    \left.\frac{d}{d\tau}\,\varphi(\tau|\varphi_\pm)\,\right|_{\;\tau=\tau_+},\\
    &&\frac{\partial^2{\mbox{\boldmath${S}$}}}
    {\partial\varphi_+\partial\varphi_-}=\dot u(\tau_+),\\
    &&u(\tau)=\frac{\partial\varphi(\tau|\varphi_\pm)}{\partial\varphi_-},
    \end{eqnarray}
where $u(\tau)$ is a particular basis function of the operator ${\mbox{\boldmath${F}$}}$ with the special boundary conditions (which follow from obvious boundary conditions for the classical solution with $\varphi(\tau_+|\varphi_\pm)=\varphi_+$, $\varphi(\tau_-|\varphi_\pm)=\varphi_-$)
    \begin{eqnarray}
    &&{\mbox{\boldmath${F}$}}u(\tau)=0,\\
    &&u(\tau_+)=0,\quad u(\tau_-)=1.
    \end{eqnarray}
This basis function reads
    \begin{eqnarray}
    u(\tau)=g(\tau)\int\limits_{\tau_+}^{\tau}
     \frac{dy}{g^2(y)}\left(g(\tau_-)\int_{\tau_+}^{\tau_-}
     \frac{dx}{g^2(x)}\right)^{-1},
    \end{eqnarray}
so that Eq.(\ref{PvVM}) gives
    \begin{eqnarray}
    {\rm Det}_{D[\tau_-,\tau_+]}{\mbox{\boldmath${F}$}}
    =C(\tau_\pm)\times g(\tau_+)\, g(\tau_-)\int\limits_{\tau_-}^{\tau_+}
    \frac{dy}{g^2(y)}                       \label{Dirichlet}
    \end{eqnarray}
where we denoted the normalization coefficient as an unknown function $C(\tau_\pm)$. In what follows we will simplify notations and, without loss of generality, put $\tau_-=0$, $\tau_+=T$, $C(\tau_\pm)\equiv C(T)$ and look for the dependence on a single parameter $T$.

Now we go over to the zeta-function calculation of the functional determinant. The determinant of the nondegenerate operator as a product of its eigenvalues,
    \begin{eqnarray}
    &&{\rm Det}_{\cal D}\,{\mbox{\boldmath${F}$}}
    =\prod_{\lambda}\lambda,             \label{prodlambda1}\\
    &&\mbox{\boldmath${F}$}
    \varphi_\lambda(\tau)
    =\lambda\,\varphi_\lambda(\tau),\quad \varphi_\lambda(\tau_\pm)=0,
    \quad \tau\in{\cal D},               \label{eigenproblem1}
    \end{eqnarray}
can be expressed in terms of the derivative of the generalized zeta-function which is just the functional trace of the inverse $s$-th power of this operator \cite{zeta}
    \begin{eqnarray}
    &&\ln {\rm Det}_{\cal D}\,
    {\mbox{\boldmath${F}$}}=
    \sum_{\lambda}\ln\lambda=-\zeta'(0),  \label{zetaprime}\\
    &&\zeta(s)=\sum_{\lambda}
    \frac1{\lambda^s}={\rm Tr}\frac1{\mbox{\boldmath${F}$}^s}=\int_{\cal D} d\tau\,\frac1{\mbox{\boldmath${F}$}^s}\,
    \delta(\tau-\tau')\,\Big|_{\;\tau'=\tau}.
    \end{eqnarray}
This function is well defined for sufficiently high positive $s$ and exists as analytic continuation in the vicinity of $s=0$.

Under the rescaling of the operator or its eigenvalues by a constant coefficient $\sigma$ the functional determinant transforms as
    \begin{eqnarray}
    {\rm Det}_{\cal D}\,\big(\sigma^2{\mbox{\boldmath${F}$}}\big) =\prod_{\lambda}\big(\sigma^2\lambda\big)=\sigma^{2\zeta(0)}
    {\rm Det}_{\cal D}\,{\mbox{\boldmath${F}$}},
    \end{eqnarray}
where $\zeta(0)$ plays the role of the number of eigenmodes of ${\mbox{\boldmath${F}$}}$ regulated by the zeta-function method \cite{anomaly}. Now we use this property to derive the $T$-dependence of ${\rm Det}_{\cal D}\,{\mbox{\boldmath${F}$}}$.

For this purpose introduce the new operator $\mbox{\boldmath${F}$}_\sigma$ of the form (\ref{operator}) with the new function $g_\sigma(\tau)$ defined on a rescaled time domain ${\cal D}_\sigma$,
    \begin{eqnarray}
    &&\mbox{\boldmath${F}$}_\sigma
    =-\frac{d^2}{d\tau^2}+\frac{\ddot g_\sigma(\tau)}{g_\sigma(\tau)},\quad \tau\in\big[0,T/\sigma\big]
    \equiv{\cal D}_\sigma,\quad
    g_\sigma(\tau)\equiv g(\sigma\tau).    \label{Fsigma}
    \end{eqnarray}
Since $\ddot g_\sigma(\tau)\equiv \frac{d^2}{d\tau^2}\,g(\sigma\tau)=\sigma^2\ddot g(\sigma\tau)$ this operator reads
    \begin{eqnarray}
    &&\mbox{\boldmath${F}$}_\sigma
    =\sigma^2\left[-\frac{d^2}{\sigma^2 d\tau^2}+\frac{\ddot g(\sigma\tau)}{g(\sigma\tau)}\,\right]=\sigma^2
    \left[-\frac{d^2}{dy^2}+\frac{\ddot g(y)}{g(y)}\,\right],\quad y=\sigma\tau\in\big[0,T\big]
    \end{eqnarray}
and has as a spectrum the set of eigenfunctions $\varphi_{\lambda_\sigma}^{(\sigma)}(\tau)$ on $\cal D_\sigma$, satisfying the same Dirichlet boundary conditions and related to the original spectrum of (\ref{prodlambda1})-(\ref{eigenproblem1}) by
    \begin{eqnarray}
    &&\varphi_{\lambda_\sigma}^{(\sigma)}(\tau)
    =\varphi_\lambda(\sigma\tau),\quad
    \lambda_\sigma=\sigma^2\lambda, \\
    &&\mbox{\boldmath${F}$}_\sigma
    \varphi_{\lambda_\sigma}^{(\sigma)}(\tau)=
    \lambda_\sigma
    \varphi_{\lambda_\sigma}^{(\sigma)}(\tau),\quad \tau\in{\cal D}_\sigma.
    \end{eqnarray}
Therefore ${\rm Det}_{\cal D_\sigma}\mbox{\boldmath${F}$}_\sigma\equiv \prod_{\lambda_\sigma}\lambda_\sigma$ equals
    \begin{eqnarray}
    {\rm Det}_{D[0,T/\sigma]}
    {\mbox{\boldmath${F}$}}_\sigma=
    \sigma^{2\zeta(0)}{\rm Det}_{D[0,T]}
    {\mbox{\boldmath${F}$}}.               \label{detrescaling}
    \end{eqnarray}

Then, applying (\ref{Dirichlet}) to the left hand side of this relation
    \begin{eqnarray}
    C(T/\sigma)\,g_\sigma(T/\sigma)\, g_\sigma(0)\int\limits_0^{T/\sigma}\frac{dy}{g_\sigma^2(y)}=
    \frac{C(T/\sigma)}\sigma\,g(T)\,g(0)\int\limits_0^T\frac{dy}{g^2(y)}
    \end{eqnarray}
and doing the same with the right hand side we obtain the equation for the coefficient function $C(T)$
    \begin{eqnarray}
    C(T/\sigma)=\sigma^{1+2\zeta(0)}C(T),  \label{Cequation}
    \end{eqnarray}
which implies that $C(T)={\rm const}\times T^{-1-2\zeta(0)}$.

To find $\zeta(0)$ we can use the heat kernel method in the approximation of the inverse mass (or proper time) expansion with the regulator mass parameter $m^2$ added to the operator $\mbox{\boldmath${F}$}$. For the differential operator acting in a $d$-dimensional spacetime with coordinates $x$ it reads
    \begin{eqnarray}
    &&\zeta_m(s)={\rm Tr}\,\frac1{\big({\mbox{\boldmath${F}$}}+m^2\big)^s}
     = \frac1{\varGamma(s)}\int\limits_0^\infty
     dt\,t^{s-1}\,{\rm Tr}\,\exp\Big\{\!-t(\mbox{\boldmath${F}$}+m^2)\Big\}
     \nonumber\\
     &&\qquad\qquad\qquad= \frac1{(4\pi)^{d/2}}\,
     \sum\limits_{n=0}^{\infty}\frac{
     \varGamma\left(s+\frac{n-d}2\right)}{\varGamma(s)}\,
     m^{d-n}A_n,                              \label{zetam}
     \end{eqnarray}
where we have used a well-known Schwinger-DeWitt (or Gilkey-Seely) proper-time expansion for the heat kernel trace \cite{DW,Gilkey-Seely,PhysRep,Avramidi,Vassilevich}
    \begin{eqnarray}
    &&{\rm Tr}\,\exp\Big\{\!-t(\mbox{\boldmath${F}$}+m^2)\Big\}
     = \frac1{(4\pi t)^{d/2}}\, e^{-t m^2}
     \sum\limits_{n=0}^{\infty}t^{n/2} A_n.
     \end{eqnarray}
Here $A_n$ represent local integrals over spacetime domain $\bf B$ and its boundaries ${\bf b}=\partial{\bf B}$, which are built of coefficients of the differential operator $\mbox{\boldmath${F}$}$. Important property of these coefficients is that for odd $n$ they are given exclusively by surface terms and, therefore, identically vanishing for problems with a compact spacetime without boundary.  For the operator of the form
    \begin{eqnarray}
    {\mbox{\boldmath${F}$}}=-\Box+P\label{F}, \quad \Box=g^{\mu\nu}\nabla_\mu\nabla_\nu,
     \end{eqnarray}
with the covariant d'Alembertian $\Box$ acting in the $d$-dimensional spacetime with coordinates $x=x^\mu$ and a generic potential term $P(x)$, the first two lowest order coefficients in the {\em Dirichlet} case are
    \begin{eqnarray}
    &A_0&= \int\limits_{\bf B} d^d\!x \,\sqrt{g} \;,
    \nonumber\\
    &A_1
      &= -\frac{\sqrt{\pi}}2 \int\limits_{\bf b} d^{d-1}\!x \,
      \sqrt{g^{(d-1)}},
        \end{eqnarray}
where the Riemann integration measures are built with respect to $d$-dimensional metric and $(d-1)$-dimensional induced metric of the boundary.\footnote{In the case of Neumann boundary conditions $A_1$ has an opposite sign on relevant boundaries.} For our case of interest $d=1$ the latter reduces to a trivial contribution of zero-dimensional points, $\int_{\bf b} d^{d-1}\!x \,\sqrt{g^{(d-1)}}=2\times1$, where the coefficient 2 signifies two end points of ${\bf B}\equiv{\cal D}=[0,T]$.

From (\ref{zetam}) the value of $\zeta_m(0)$, which is always given by the first $d$ terms of the local Schwinger-DeWitt expansion (analytic in $m$ with $n\leq d$), in the case of $d=1$ reduces to the boundary term $A_1$ and reads
    \begin{eqnarray}
    \zeta(0)=\zeta_m(0)\,\Big|_{\;m\to 0}=\frac{A_1}{\sqrt{4\pi}}=-\frac12.
     \end{eqnarray}
Therefore $1+2\zeta(0)=0$, and $C(T)$ is a constant independent of $T$, which fully confirms the Pauli-van Vleck-Morette formula (\ref{PvVM}).

\subsection{Periodic boundary conditions}

For the periodic boundary conditions with the multi-node zero mode $g(\tau)$ in the domain
    \begin{eqnarray}
    &&{\cal D}=S^1,\quad 0=\tau_0<\tau<T,\\
    &&0=\tau_0<\tau_1<...<\tau_{2k}=T,
    \end{eqnarray}
the reduced functional determinant is determined by Eqs.(\ref{prodlambda})-(\ref{eigenproblem}) which, similarly to (\ref{zetaprime}) bring us to the use of the {\em reduced} zeta-function
    \begin{eqnarray}
    &&\zeta^*(s)=
    \sum_{\lambda\neq 0}\frac1{\lambda^s}.       \label{zetastar}
    \end{eqnarray}
The scaling behavior is again determined by $\zeta^*(0)$ and reads
    \begin{eqnarray}
    {\rm Det}_{\cal D}^*\,\big(\sigma^2{\mbox{\boldmath${F}$}}\big) =\prod_{\lambda\neq 0}\big(\sigma^2\lambda\big)=\sigma^{2\zeta^*(0)}
    {\rm Det}_{\cal D}^*\,{\mbox{\boldmath${F}$}},   \label{scaling}
    \end{eqnarray}
which allows us to repeat the steps of the previous section.

Consider again the operator (\ref{Fsigma}) on a rescaled circle with the rescaled set of zeroes of $g_\sigma(\tau)$, $g_\sigma(\tau_i^\sigma)=0$,
    \begin{eqnarray}
    &&{\cal D}_\sigma=S^1_\sigma,\quad 0=\tau_0<\tau<\frac{T}\sigma,\\
    &&0=\tau_0^\sigma<\tau_1^\sigma<...
    <\tau_{2k}^\sigma=\frac{T}\sigma,\quad \tau_i^\sigma=\frac{\tau_i}\sigma.
    \end{eqnarray}
Its reduced functional determinant is given by Eq.(\ref{det1})
    \begin{eqnarray}
    &&{\rm Det}_{\cal D_\sigma}^*\,
    {\mbox{\boldmath${F}$}_\sigma}=C\left(\frac{T}\sigma\right)\,
    \Delta^\sigma\oint_{\cal D_\sigma}
    d\tau\,g^2_\sigma(\tau)=\frac1{\sigma^2}\,
    C\left(\frac{T}\sigma\right)\,
    \Delta\oint_{\cal D}
    d\tau\,g^2(\tau).                    \label{det2}
    \end{eqnarray}
Here we took into account simple expressions for the monodromy and the zero mode norm of the rescaled operator
    \begin{eqnarray}
    &&\Delta^\sigma=\frac\Delta\sigma,\\
    &&\oint_{\cal D_\sigma}
    d\tau\,g^2_\sigma(\tau)=\frac1\sigma
    \oint_{\cal D} d\tau\,g^2(\tau),
    \end{eqnarray}
which in their turn follow from the following obvious relations for its partial basis functions $\varPsi_i^\sigma(\tau)$ and its monodromy constituents $\Delta_i^\sigma$ in $\Delta^\sigma=\sum_i \Delta^\sigma_i$,
    \begin{eqnarray}
     &&\varPsi_{i}^\sigma(\tau)= g_\sigma(\tau)
     \int\limits_{\tau_{i}^*/\sigma}^{\tau}
     \frac{dy}{g^2_\sigma(y)}=\frac1\sigma\,
     \varPsi_{i}(\sigma\tau),\;
     \quad
     \tau_{i-1}^\sigma<\tau<\tau_i^\sigma\;,
     \quad i=1,...,2k,                         \\
     &&\Delta^\sigma_i=-\varPsi_{i}^\sigma
     \dot\varPsi_{i}^\sigma
     \Big|_{\;\tau_{i-1}^\sigma}^{\;\tau_i^\sigma}
     =\frac{\Delta_i}\sigma.
    \end{eqnarray}

Applying (\ref{det2}) and (\ref{det1}) respectively to the left-hand and right-hand sides of Eq.(\ref{scaling}) we get the equation for $C(T)$ analogous to (\ref{Cequation}),
    \begin{eqnarray}
    C(T/\sigma)=\sigma^{2+2\zeta^*(0)}C(T).   \label{Cequation1}
    \end{eqnarray}

For the calculation of $\zeta^*(0)$ we again consider the generalized zeta-function regulated by a sufficiently large mass parameter, so that all eigenvalues $\lambda+m^2$ are positive,
    \begin{eqnarray}
    &&\zeta_m(s)=\sum\limits_\lambda
    \frac1{\big(\lambda+m^2\big)^s}=\sum\limits_{\lambda\neq 0}
    \frac1{\big(\lambda+m^2\big)^s}+\frac1{m^{2s}}\equiv
    \zeta^*_m(s)+\frac1{m^{2s}},
     \end{eqnarray}
whence
    \begin{eqnarray}
    \zeta^*_m(0)=\zeta_m(0)-1.
     \end{eqnarray}

Note that $\zeta_m(s)$ here is not the reduced one, and it contains the contribution of $\lambda=0$ shifted by a large $m^2$ to the positive range. Therefore, the heat kernel representation applies to it, and the value of $\zeta_m(0)$ is again given by Eq.(\ref{zetam}), but this time for a compact domain ${\cal D}=S^1$ without boundaries. In this case all odd number coefficients are vanishing $A_{2k+1}=0$, because they are exclusively contributed by boundary terms, and $\zeta_m(0)=0$. Therefore, $\zeta^*(0)=\zeta^*_m(0)=-1$ and the coefficient $C(T)$ in (\ref{Cequation1}) and (\ref{det1}) is again a constant independent of $T$.

Its actual value within zeta-function regularization can be determined for a particular case of the constant function $g(\tau)=c$ corresponding to the operator $\mbox{\boldmath${F}$}=-d^2/d\tau^2$ with the explicit spectrum of eigenfunctions and respective eigenvalues
    \begin{eqnarray}
    &&\varphi_0=1,\quad \lambda_0=0,\\
    &&\varphi_{1\,n}(\tau)=\sin\left(\frac{2\pi n}{T}\,\tau\right),\; \varphi_{2\,n}(\tau)=\cos\left(\frac{2\pi m}{T}\,\tau\right),
    \quad \lambda_n=\left(\frac{2\pi n}{T}\right)^2,
    \quad n=1,2,...\;.
    \end{eqnarray}
The logarithm of the corresponding restricted determinant -- the product of all nonvanishing eigenvalues regularized by zeta-function method -- equals
    \begin{eqnarray}
    \ln{\rm Det_*}\!\left(-\frac{d^2}{d\tau^2}\right)
    =2\sum\limits_{n=1}^\infty
    \ln\left(\frac{2\pi n}{T}\right)^2
    =4\ln\left(\frac{2\pi}{T}\right)
    \zeta_R(0)-4\zeta'_R(0)=2\ln T.   \label{1000}
    \end{eqnarray}
Here $\zeta_R(s)=\sum_{n=1}^\infty n^{-s}$ is the Riemann zeta function having the following particular value $\zeta_R(0)=-\frac12$ and the value of its derivative $\zeta_R'(0)=-\frac12\ln{2\pi}$. On the other hand, the basis functions and the monodromy $\Delta$ for this operator read
    \begin{eqnarray}
    &&g(\tau)=c,\quad \oint d\tau\,g^2=c^2\,T,\nonumber\\
    &&\psi(\tau)=\frac1c(\tau-\tau_*),\quad\Delta=
    \frac{T}{c^2}.
    \end{eqnarray}
Therefore according to (\ref{det1}) ${\rm Det_*}\!\left(-d^2/d\tau^2\right)=C(T)\,T^2$, and the comparison with (\ref{1000}) gives the $T$-independent result $C(T)=1$ for the normalization coefficient in (\ref{det1}).

\section{Conclusions}
Thus we see that the above combination of methods gives exhaustive answer for the reduced functional determinant of the operator (\ref{operator}) having a multi-node zero mode in the periodic boundary value problem. This determinant expresses in terms of the monodromy of its basis function, which is obtained in quadratures as a sum of contributions (\ref{Delta})-(\ref{Deltai}) of time segments connecting neighboring pairs of the zero mode roots within the period range. Few words are in order here just to reiterate our special interest in this particular problem, briefly mentioned in Introduction.

The operator $\mbox{\boldmath${F}$}$ determines the one-loop statistical sum for the microcanonical ensemble in cosmology generated by a conformal field theory \cite{slih,why,PIQC}. This ensemble realizes the concept of cosmological initial conditions by generalizing the notion of the no-boundary wavefunction of the Universe to the level of a special quasi-thermal state which is dominated by instantons with an oscillating cosmological scale factor $a(\tau)$ of their Euclidean FRW metric. These oscillations result in the multi-node nature of the zero mode $g(\tau)\sim \dot a(\tau)$ of $\mbox{\boldmath${F}$}$, which itself arises as the residual conformal Killing symmetry of the FRW background. This, in particular, explains the motivation for the gauge-fixing treatment of the zero mode considered above.

As was mentioned above, a very attractive feature of the cosmological microcanonical ensemble is that in the case of the CFT driven cosmology it suggests a possible solution of the cosmological constant problem -- the restriction of the range of the primordial $\Lambda$ by a new quantum gravity scale, its value being encoded in the conformal anomaly of the underlying CFT \cite{slih,why}. Moreover, as suggested in \cite{DGP/CFT}, these microcanonical initial conditions admit inflationary scenario in the early Universe and can provide a thermal input in the red tilt of the COBE part of the CMB spectrum. This tilt can be additional or, perhaps, even alternative to the conventional red tilt generated from primordial vacuum fluctuations of \cite{MukhanovChibisov}. This makes the hypothesis of microcanonical initial conditions in quantum cosmology not only feasible, but also observationally verifiable, perhaps, in a foreseeable future. Also, the statistical sum of this ensemble is likely to predict interesting phase transitions for multi-node cosmological instantons \cite{oneloop1} which makes physics of this model very rich and interesting. The results and methods presented above seem indispensable for a further progress in these intriguing issues.

\section*{Acknowledgments}
The authors are grateful to A.Yu.Kamenshchik, I.V.Tyutin and B.L.Voronov for helpful discussions. The work  A.B. was supported by the RFBR grant No. 11-01-00830 and D.N. was supported by the RFBR grant No. 11-02-00512.

\end{document}